\documentclass[11pt]{article}
\usepackage{blois,epsfig}

\def\apj{{\em ApJ}}
\def\mnras{{\em MNRAS}}

\def\be{\begin{equation}}
\def\ee{\end{equation}}
\def\bea{\begin{eqnarray}}
\def\eea{\end{eqnarray}}

\begin{document}
\vspace*{4cm}
\title{GALAXY EVOLUTION FROM THE DEEP2 GALAXY REDSHIFT SURVEY}

\author{ D.S. MADGWICK\footnote{On behalf of the DEEP2 Team: 
Alison L. Coil, 
Christopher J. Conselice, 
Michael C. Cooper, 
Marc Davis (P.I), 
Richard S. Ellis, 
Sandra M. Faber, 
Douglas P. Finkbeiner, 
Brian F. Gerke, 
Puragra Guhathakurta, 
Nick Kaiser,
David C. Koo, 
Jeffrey A. Newman, 
Andrew C. Phillips, 
Charles C. Steidel, 
Benjamin J. Weiner and 
Christopher N. A. Willmer 
}}

\address{Department of Astronomy, Campbell Hall, U.C. Berkeley,\\
Berkeley CA94720, USA}

\maketitle\abstracts{
The DEEP2 Galaxy Redshift Survey is now well underway, having already measured the redshifts to $\sim$5600 galaxies in its first season of observations.  Here I briefly review the survey itself, before discussing in more detail some of the initial science results to have recently appeared in the literature.  In particular, the potential of the survey to characterize galaxy evolution is discussed, with special emphasis on the role of spectral classification.  Some of the applications of this classification, namely to the quantification of galaxy clustering and the measurement of galaxy environments are also discussed here.
}

\section{The DEEP2 Survey}

The DEEP2 (DEEP Extragalactic Evolutionary Probe 2) Redshift Survey (Davis et al.\cite{dav02}) is an ambitious project to quantify the galaxy population, and in particular it's large-scale structure, at $z\sim1$.  In so doing the survey will not only reveal the detailed properties of high-redshift galaxies, but will also provide a valuable cosmological probe, particularly when combined with such large local surveys as the 2dFGRS or SDSS (Colless et al.\cite{col01}; Strauss et al.\cite{strau02}). 

In order to achieve its goals, the survey makes use of the advanced multiplexing ability and high spectral resolution of the recently commissioned DEIMOS instrument on the Keck-II telescope at Hawaii.  In so doing, the survey is able to routinely determine highly accurate redshifts, and should be able to easily achieve its goal of 60,000 redshifts in only 120 nights over a three year period.  Such an achievement would represent an order of magnitude improvement over any previous study of such distant galaxies.  

In its first season of observations (August -- October, 2003), the
DEEP2 Redshift Survey has already accurately 
measured the redshifts of $\sim5600$ galaxies, out of this proposed total
of 60,000, and it is the applications of this initial data-set that will be focussed on in these proceedings. 

\subsection{Galaxy selection}

The galaxies observed as part of the DEEP2 Redshift Survey occupy 
four independent fields, for which extensive $B$,$R$ and $I$ CFHT 12k x 8k 
photometry is available.  Each of these fields is roughly 2 x 0.5 deg$^2$ on the sky, corresponding to comoving dimensions of 80 x 20 $h^{-1}$Mpc at $z=1$ (assuming a $\Lambda$CDM Universe).  In addition to this, the line-of-sight dimension of the survey  (from $z=0.7-1.4$) corresponds to $\sim1400$ $h^{-1}$ Mpc, so that the survey is very much optimized towards evolutionary studies over a broad range of redshifts.  Each of the fields has been chosen to occupy low-extinction regions that are continuously observable from Hawaii over a six month interval.  In addition, one of the fields has been chosen to overlap the extended Groth Survey strip (Groth et al.\cite{groth}).

\begin{figure}
\begin{center}
\epsfig{file=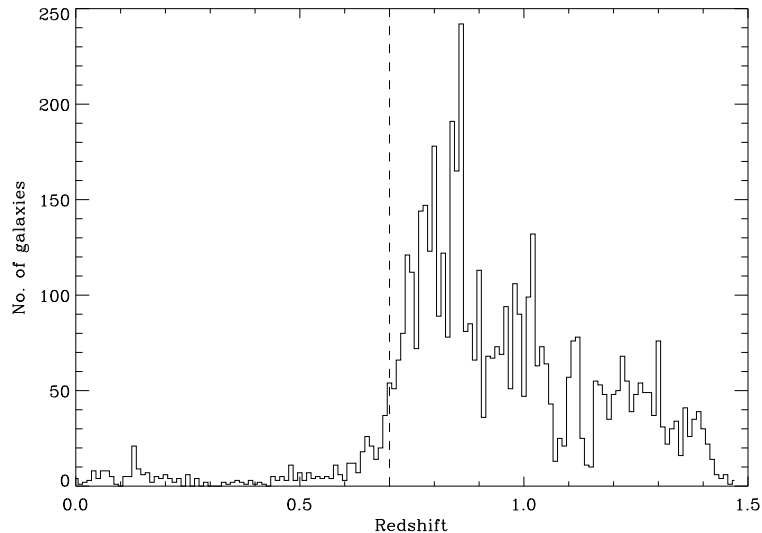,height=3in}
\caption{The redshift distribution of the galaxies observed so far in the DEEP2 Galaxy Redshift Survey.  It is clear from this figure that the simple color-based preselection of high-redshift galaxies is working very well, and is consistently selecting only galaxies with $z>0.7$.  
\label{fig:color}}
\end{center}
\end{figure}

Galaxies are selected from the available photometry to an extinction corrected limit
$R_{\rm AB}<24.1$, and, in all but the Groth field, a simple color-cut (based upon only the observed $R-I$ and $B-R$ colors) has been adopted to select only high-redshift galaxies ($z>0.7$).  It has been shown previously (Davis et al.\cite{dav02}) that this pre-selection of high-redshift galaxies is very effective, and this is also demonstrated in Fig.\ref{fig:color}, where the redshift distribution of the first seasons observations is shown.  Note that in the field overlapping the Groth strip we will not apply this pre-selection to the galaxies, so that the full evolutionary history from $z=0$ can be studied there.

A sophisticated automated pipeline has been developed to efficiently extract 
and reduce the spectra
observed in the survey, whilst preserving all the useful kinematic details available (such as rotation curve information).  Details of this pipeline 
will be presented in Newman et al.\cite{newman}. 
The observed spectra themselves are taken at particularly high resolution 
($R\sim5000$) using the DEIMOS spectrograph mounted on the Keck-II
telescope (Davis \& Faber\cite{dav98}), 
and generally span the wavelength range of $6400 <
\lambda < 9200$\AA.  Note that having such high-resolution spectra is crucial to our analysis, in that it allows us to easily work between the bright sky-lines that are particularly prominent over this wavelength interval.  For galaxies with $z > 0.7$ this wavelength window allows us to
measure the redshift of each galaxy, particularly when the
resolved [O{\sc ii}] doublet is present 
(out to a redshift of $z=1.5$).  Absorption based redshifts are also
readily determined, primarily using the Ca H+K features which are
visible to redshifts of $z\sim1.3$.

In total, we currently obtain redshifts for approximately 70\% of all galaxies targeted in the survey, although with recent improvements in the data-reduction pipeline this is soon expected to reach 75-80\%.

\section{Spectral classification}

The classification of galaxies is of fundamental importance for
understanding galaxy populations, and for this reason is a
very important aspect of any galaxy redshift survey.  
Having a data set of
many thousands of galaxy spectra allows one to test the
validity of galaxy 
formation and evolution scenarios with unprecedented accuracy.  However,
the sheer size of the full spectral data set presents its 
own unique problems.
In order to make such a galaxy data set more 
`digestible' some form of data
compression is necessary, whether this be through the adoption of
morphological classification, colors or some other
compression/classification scheme.
If these quantities can be determined consistently over a wide
range of redshifts, they can be compared with theoretical predictions and
simulations, and hence set constraints on scenarios for
galaxy evolution.  This will be especially true if consistent
classification regimes can be used for both high redshift ($z\sim1$)
surveys, such as 
DEEP2, and the large $z\sim 0$ surveys
now approaching full 
completion; the SDSS
and the 2dFGRS.

A number of
different approaches to the
classification of galaxy spectra have been adopted for local galaxy surveys.
These include the calculation of
rest-frame colors (e.g. Strateva et al.\cite{stra01});
principal component analysis (PCA) based spectral classifications
(e.g. Connolly et al.\cite{con95}; Bromley et al.\cite{brom98}; Folkes et al.\cite{folk99};
Madgwick et al.\cite{mad02}; de Lapparent et al.\cite{lap03}), and other more sophisticated discriminations
(e.g. Heavens, Jimenez \& Lahav\cite{heav00}; Slonim et al.\cite{slon00}), based upon
information theory.  The
underlying theme of all these alternative methods is that they
characterize the galaxy population exclusively in terms
of their observed spectra.  

\subsection{Classifying galaxies at $z\sim1$}

The DEEP2 Galaxy Redshift Survey has implemented and tested the use of a PCA-based spectral classification for galaxy classification (Madgwick et al.\cite{mad04}).
In so doing we have presented a new spectral
classification, $\eta_{\rm DEEP}$, 
for the galaxies observed to date in the DEEP2 Redshift 
Survey.  The main goal in developing this classification was to
provide a consistent and robust measure of the type of a galaxy
over the large range of redshifts encountered in this survey.
To do this special handling of incomplete and `gappy' observed spectra
is required.

Based upon previous analyses (Madgwick et al.\cite{mad03}), there is strong evidence
that this parameter, $\eta_{\rm DEEP}$, will
correlate well with
the relative amount of star formation as expressed by the Scalo birthrate parameter (Scalo\cite{scal86}),
\begin{equation}
b_{\rm Scalo} = \frac{SFR_{\rm current}}{\langle SFR\rangle_{\rm past}} \;.
\end{equation}
A more detailed study of this correlation, together with the
role of higher order PCA projections, will be forthcoming when higher
resolution spectral synthesis models become publicly available in the
near future (the DEEP2 galaxy spectra are at a resolution $\sim20$
times higher than any publicly available synthesis model!).

\begin{figure}
\begin{center}
\epsfig{file=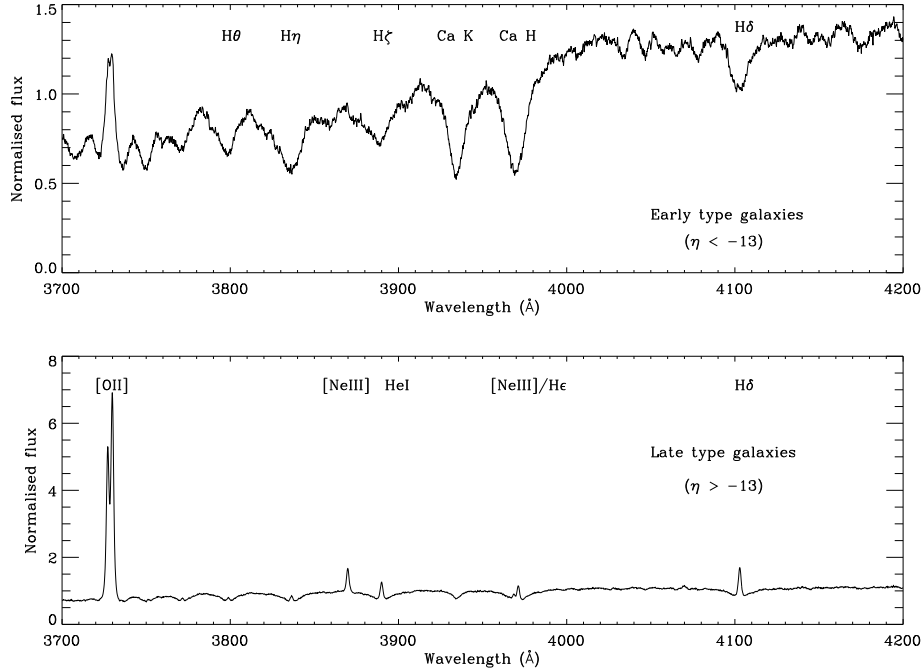,height=5in,angle=90}
\caption{The average spectra of our two different spectral types is shown.  The different relative prominence of the nebular emission features is readily visible from this plot, showing that the classification is readily separating those galaxies with relatively more recent star-formation activity from their more passive counterparts.
\label{fig:spec}}
\end{center}
\end{figure}

This classification, $\eta$, will also be particularly
useful in subsequent analyses of the galaxy luminosity function
or correlation functions, as it is easily comparable with other
classifications at $z=0$, for which large spectroscopic samples are
now publicly available.  In addition, previous work (e.g. Madgwick
et al.\cite{mad03}) has shown that it is straightforward to make direct
comparisons between such a spectral classification regime and the
output of semi-analytic galaxy models (e.g. Kauffmann, White \&
Guiderdoni\cite{kauf93}; Cole et al.\cite{cole94}; Somerville \& Primack\cite{somer99}),
allowing us to directly constrain the assumed models of 
galaxy formation and evolution between $z=1$ and $z=0$ using this form
of classification.

\section{Galaxy clustering}

It is now well established that the clustering of galaxies at low
redshift depends upon a variety of factors.  Perhaps one of the most potentially interesting of these, which has been discussed extensively in the literature, is the variation in the observed clustering with the
galaxy type under consideration (e.g. Davis \& Geller 1976; Dressler\cite{dres80}; Lahav, Nemiroff
\& Piran\cite{lnp}; Hermit et al.\cite{herm96}; Madgwick et al.\cite{madxi}).  
Of particular interest then, is to establish whether the same trends are also observable at much higher redshifts, and what constraints these can place on the evolution processes present over this period.
Towards this goal, an initial quantification of the two-point correlation function has recently been carried out using the first season's data (see Coil et al.\cite{coil}) from the DEEP2 Redshift Survey.

\begin{figure}
\begin{center}
\epsfig{file=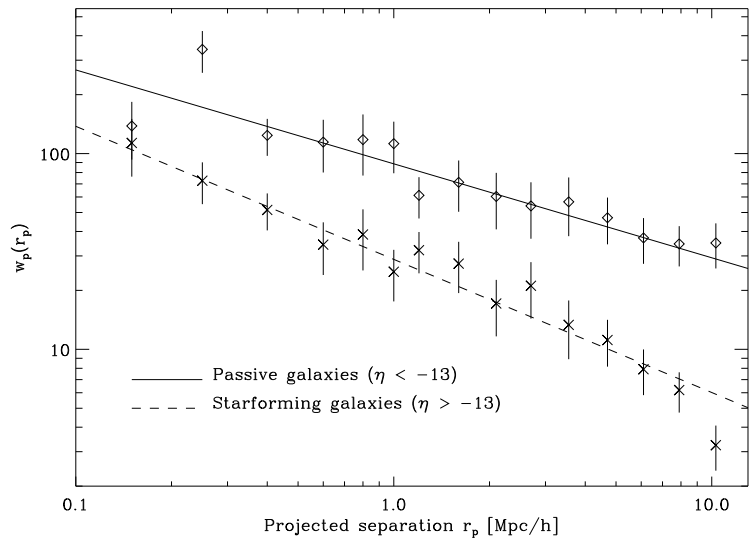,height=3in}
\caption{The projected correlation functions are shown for different types of galaxies in the DEEP2 Redshift Survey (Coil et al. 2003).  It can be seen from this figure that galaxies undergoing very little star-formation ($\eta<-13$) are much more strongly clustered than those with significant amounts of recent star-formation ($r_0=6.6\pm1.1\;h^{-1}$ Mpc and $r_0=3.2\pm0.5\;h^{-1}$ Mpc respectively). 
\label{fig:xi}}
\end{center}
\end{figure}

The analysis presented by Coil et al.\cite{coil} made use of only 2219 galaxies drawn from a contiguous region of the initial observations and hence only represents $\sim5$\% of the total potential of the survey.  However, despite this, very useful constraints were able to be made as to the relative normalization of the two-point correlation function.  In particular it was noted that the observed clustering of the most actively star-forming galaxies in the sample was significantly lower than that of the less active galaxies, perhaps even more so than what is observed locally (see Fig.~\ref{fig:xi}).

As the size of the available data-set continues to increase it will be interesting to extend this analysis to investigate in particular the shape of these correlation functions (as opposed to just their normalization's) as this can yield a great deal of useful information, for example about the processes governing the galaxy population under consideration (see Yan, Madgwick \& White~\cite{yan} for further discussion).  In addition, as the data-set increases it will soon become possible to `de-couple' the effects of various other properties of the samples under consideration, so that more detailed comparisons can be made (e.g. Norberg et al.\cite{nor}).

\section{Galaxy environments}

The observed properties of the galaxy population have long been known
to depend  
upon the environment in which they are located.  For example,
red, non-starforming galaxies (such as elliptical and lenticular
galaxies) are found to be systematically over-represented in dense
environments -- such as groups and clusters (e.g. Dressler\cite{dres80};
Balogh et al.\cite{bal97}; Martinez et al.\cite{mart02}).  

Presently there are a variety of mechanisms that can readily
explain these observational trends; such as the action of dynamical
friction, tidal stripping or gas pressure in densely occupied
environments.  These mechanisms, together with the hierarchical model of
galaxy formation -- in which galaxies formed in less
dense environments and are then accreted into larger groups and clusters -- is
generally consistent with the present observations.  However, it is
presently unclear to what extent the environment of a galaxy affects
its subsequent evolution, since the environment in which a galaxy has formed
may be substantially different to the one it inhabits in low-redshift
galaxy surveys (Carlberg\cite{carl}).  

The determination of the exact influence of galaxy environment upon 
evolution is one area where the advent of
DEEP2 will be able to significantly advance our
understanding of the processes involved in galaxy evolution -- by giving
us a representative snap-shot of the galaxy population and it's
environments when the Universe was half its present age.  In particular, 
when this
survey is combined with those currently available at low-redshift (such 
as the 2dFGRS  or the SDSS), we will be able to build a comprehensive picture of the 
evolution of the galaxy population with which detailed models can be 
readily tested.

\subsection{Defining environment}

The `environment' of a galaxy is usually defined in terms of the density 
of galaxies in its immediate vicinity.  For example, previous 
analyses have focussed on the identification of  galaxies in large groups or 
clusters, which can be contrasted to those galaxies not 
inhabiting these over-dense regions (the field population).  
Another approach is to instead derive a continuous measure of the galaxy density distribution, e.g. by measuring the distance to the $n$$^{\rm th}$-nearest neighbor (e.g. Gomez et al.\cite{gomez}), or by directly smoothing the distribution on a fixed scale (e.g. Hogg et al.\cite{hogg}).

The underlying theme in all of these methods is that one requires a measure of the local galaxy number density {\em around} each galaxy.  Based upon this we have chosen to define the environment of galaxies in the DEEP2 Redshift Survey
slightly differently, by instead using the unique volume inhabited by each galaxy (which is obtained 
using a voronoi tessellation of the data set). In so doing we can define 
a continuous sequence of (inverse) galaxy density, without recourse to group 
definitions, or the need to smooth the galaxy distribution in any way. The main difference between this method of measuring local density and those used previously, is that the voronoi volume
is asymptotically local: the density measured
at a given point is wholly determined by the neighboring data points,
with the influence of distant points vanishing entirely.

A voronoi polyhedron is the unique three-dimensional convex
region of space surrounding a data point (in this case a galaxy), within which every point
is closer to the galaxy than to any other data point (see e.g. Marinoni et al.\cite{mar02}).  
The faces of the
Voronoi polyhedron are defined by the perpendicular bisecting planes
of the vectors connecting the seed to its neighbors.  A galaxy's
neighbors are those points connected to it by the Delaunay
complex---the set of tetrahedra whose vertices are at the data
points and whose unique, circumscribing spheres contain no other data
points.

The voronoi partition and Delaunay complex also contain much information
about the spatial clustering properties of galaxies.
Indeed, we initially computed the Voronoi partition for the DEEP2
data-set in the context of the Voronoi-Delaunay group-finding method
(VDM) of Marinoni et al.\cite{mar02}.   Our methods for computing the
Voronoi partition are identical to theirs, and we refer the reader to
that work for computational details and for further discussion
of the usefulness and historical context of the Voronoi partition and
Delaunay complex.  A catalogue of DEEP2 groups identified using the
VDM will be presented in Gerke et al.\cite{gerk03}

\begin{figure}
\begin{center}
\epsfig{file=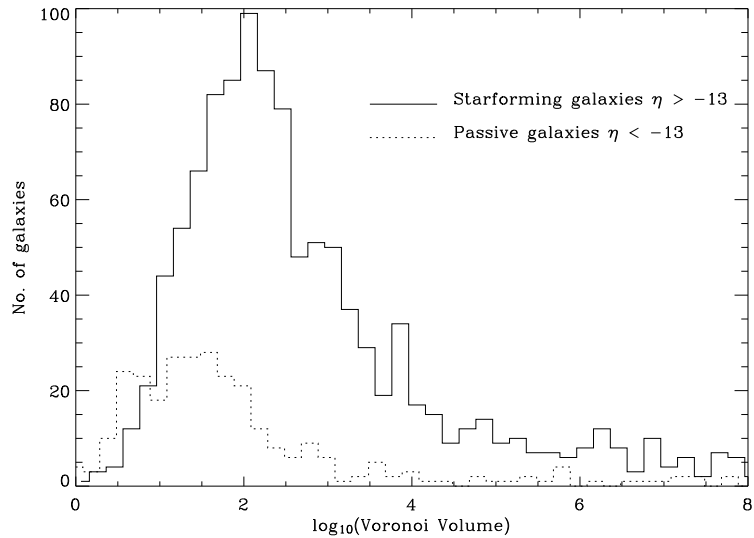,height=3in}
\caption{The different environments (defined in terms of the unique volume occupied by each galaxy) for different types of galaxies is shown here.  The x-axis corresponds to the volume of a voronoi cell that has been constructed around each galaxy and is inversely proportional to the local galaxy number density.  It can be seen from this figure that galaxies undergoing very little star-formation ($\eta<-13$) have predominantly small volumes -- corresponding to dense environments.
\label{fig:env}}
\end{center}
\end{figure}

The voronoi volume distribution of different types in the DEEP2 Survey (as defined by the $\eta$ spectral type parameter) is shown in
Fig.~\ref{fig:env}, in which it can be seen that the least actively
star-forming galaxies ($\eta<-13$) have systematically smaller voronoi volumes -- showing 
that they preferentially occupy the densest environments.  This result is 
consistent with previous analyses at lower redshifts (e.g. Balogh et al.\cite{bal97}), and shows that the environment has already had a significant 
impact on the presence of star-formation in galaxies as far back as $z\sim1$.
We are in the process of performing a similar analysis on the 2dFGRS and SDSS, which will allow us to easily make direct comparisons over this redshift interval.

\section{Conclusions}

The DEEP2 Galaxy Redshift Survey is now making significant progress, having already measured redshifts to $\sim$5600 galaxies.  In these proceedings I have briefly summarized a few of the recent results to have come out of these initial observations, focussing in particular on the role of spectral classification in characterizing galaxy evolution.  

As more data becomes available in the near future, it is expected that the quantification of this evolution in the galaxy population will provide particularly strong constraints on models of galaxy formation and evolution.  This, along with detailed measurements of the large-scale structure at $z\sim1$, will present significant advances in our understanding of both extra-galactic astronomy and cosmology, both of which are made possible by the remarkable scope of this survey.

\section*{Acknowledgments}

This work was supported in part by NSF grants AST00-71048
and KDI-9872979.   The DEIMOS spectrograph was funded by a grant from CARA
(Keck Observatory), by an NSF Facilities and Infrastructure grant
(AST92-2540),  
by the Center for Particle Astrophysics,
and by gifts from Sun Microsystems and the Quantum Corporation.
DPF is supported by a Hubble Fellowship.

The DEEP2 Redshift Survey has been made possible through the dedicated
efforts of the DEIMOS staff at UC Santa Cruz who built the instrument
and the Keck Observatory staff who have supported it on the telescope.

\end{document}